\documentclass[aps,prb,twocolumn,superscriptaddress,
floatfix,amsmath,amssymb,nofootinbib,longbibliography]{revtex4-2}
\usepackage[utf8]{inputenc} 
\usepackage[T1]{fontenc} 
\usepackage{mathtools}
\usepackage{float}
\usepackage{siunitx}
\usepackage{physics}
\usepackage{graphicx}
\usepackage{enumerate}
\usepackage{hyperref}
\hypersetup{
    colorlinks=true,       
    linkcolor=blue,          
    citecolor=blue,       
    filecolor=magenta,      
    urlcolor=blue          
}
\usepackage{bm}
\usepackage{soul}

\renewcommand\Re{\operatorname{Re}}
\renewcommand\Im{\operatorname{Im}}

\begin{document}

\title{Theory of infrared magneto-optical effects from chiral phonons in solids}

\author{Chiara Fiorazzo}
\email{chiara.fiorazzo@unitn.it}
\affiliation{Department of Physics, University of Trento, Via Sommarive 14, 38123 Povo, Italy}
\author{Cheol-Hwan Park}
\email{cheolhwan@snu.ac.kr}
\affiliation{Department of Physics and Astronomy, Seoul National University, Seoul 08826, Korea}
\affiliation{Center for Theoretical Physics, Seoul National University, Seoul 08826, Korea}
\affiliation{Centro de F\'isica de Materiales, Universidad del Pa\'is Vasco, 20018 San Sebasti\'an, Spain}
\affiliation{Donostia International Physics Center, 20018 San Sebasti\'an, Spain}
\author{Ivo Souza}
\email{ivo.souza@ehu.es}
\affiliation{Centro de F\'isica de Materiales, Universidad del Pa\'is Vasco, 20018 San Sebasti\'an, Spain}
\affiliation{Ikerbasque Foundation, 48013 Bilbao, Spain}
\author{Matteo Calandra}
\email{m.calandrabuonaura@unitn.it}
\affiliation{Department of Physics, University of Trento, Via Sommarive 14, 38123 Povo, Italy}


\date{\today}

\begin{abstract}
In crystals with broken time-reversal symmetry,
zone-center phonons can acquire a finite angular momentum via
velocity-dependent  forces on the nuclei.  Despite
having the same order of magnitude as the electron spin angular
momentum, the phonon angular momentum can be hard to detect because
the frequency splitting is small.
Here, by developing a theory of lattice magneto-optical effects in
reflection and transmission, we show that infrared magnetic circular
dichroism is a sensitive probe of zone-center phonon chirality.
%
%
We evaluate the infrared magneto-optical Faraday, Kerr, and
circular-dichroism spectra of CrI$_3$
from time-dependent density-functional theory in the adiabatic
local-density approximation.
We find  sizeable 
circular dichroism from the infrared-active E$_u$ mode at $\approx 214$ cm$^{-1}$, even though the calculated splitting is
only 0.22 cm$^{-1}$.
\end{abstract}
\maketitle
\section{Introduction}

The observation of the phonon Hall effect in
2005~\cite{PHE_experimental} brought renewed attention to the subtle coupling
of lattice vibrations to external magnetic fields~\cite{MeadRevModPhys.64.51,resta-jpcm00} and, by extension,
to internal magnetic
order~\cite{qin-prl12,Zhang_NiuPhysRevLett.112.085503,saito-prl19,saparov-prb22}. 
Lattice vibrations are typically
treated in the Born-Oppenheimer approximation~\cite{born-book54},
where the only (harmonic) coupling to electrons is via the interatomic
force-constant (IFC) matrix.  Since the IFC is a static response
function, any time-reversal breaking that may be present in the
electronic wave functions is not transmitted to the lattice degrees of
freedom. As a result, phonon Hall and other effects arising from
broken time-reversal phonons are absent from that description; in
particular, nondegenerate zone-center phonon modes are linearly
polarized (away from $\Gamma$ they can become elliptically polarized
if inversion symmetry is broken, leading to observable
effects~\cite{OPTICAL_Observation_of_chiral_phonons,Edelstein}).
  %
%
%

Once time-reversal is broken in the lattice dynamics, degenerate
linearly-polarized phonon modes at $\Gamma$ may split into
nondegenerate elliptically-polarized (``chiral'') modes carrying a net
angular momentum~\cite{Zhang_NiuPhysRevLett.112.085503}.  To capture
such effect, one must go beyond the static approximation in the
description of interatomic forces.  In the Born-Oppenheimer scheme,
this is achieved by including the nuclear geometric vector
potential~\cite{MeadTruhlar10.1063/1.437734,MeadRevModPhys.64.51},
which gives rise to velocity-dependent forces mediated by the nuclear
Berry curvature ~\cite{resta-jpcm00,saito-prl19} . Alternatively, one may start
from an nonadiabatic (dynamical) electron-phonon response function;
the ``velocity-force'' coupling is then recovered by expanding the
dynamical IFC matrix to first order in
frequency~\cite{BistoniPhysRevLett.126.225703}.

Recently, chiral vibration modes induced by velocity-dependent forces
have been studied from first principles in magnetic
molecules~\cite{BistoniPhysRevLett.126.225703} and
solids~\cite{BoniniPhysRevLett.130.086701,RenPhysRevX.14.011041}.
These works highlighted the role of relativistic effects and
noncollinear magnetism in the generation of a net vibrational angular
momentum.  
Experimental detection of this effect is a challenging task; the main
difficulty is that the frequency splittings
are small, of the order of $1$ cm$^{-1}$ or less, even when the phonon
angular momentum is of the order of $\hbar/2$.  Nevertheless, the
chirality of Raman-active zone-center phonons in magnetic crystals has
been detected by the reversal of helicity of circularly-polarized
incident light~\cite{du-advmater19,yin-advmater21}, and possible
signatures in infrared (IR) spectroscopy have also been
suggested~\cite{BistoniPhysRevLett.126.225703,RenPhysRevX.14.011041}.


In this work, we propose IR magneto-optical effects as sensitive
probes of mode-resolved zone-center phonon angular momentum.  Quite generally,
magneto-optical effects in the electric-dipole approximation (Faraday
and Kerr rotation, magnetic circular dichroism\cite{shinagawa00}) are
sensitive to a net (macroscopic) magnetization. They are associated
with the presence of a nonzero antisymmetric (AS) part in the
dielectric function,
\begin{equation}
\epsilon_{\alpha\beta}^{\rm AS} (\omega)=
\frac{\epsilon_{\alpha\beta}(\omega)-\epsilon_{\beta\alpha}(\omega)}{2},
\label{eq:epsil_AS_def}
\end{equation}
which is time-odd according to the
Onsager relation.
Working in the framework of time-dependent density-functional theory
(TDDFT), we evaluate the lattice-mediated 
contribution to Eq.~\eqref{eq:epsil_AS_def}  in the adiabatic
local density approximation find that
  it is induced by the  nuclear Berry
curvature.  In addition, we obtain expressions for the magnetic
circular dichroic absorbance and for the Faraday and Kerr rotation
strengths, taking into account the sample thickness for effects in
transmission.
%
%
We apply our formalism to bulk CrI$_3$ and show that the predicted
circular dichroism is sizeable, suggesting that it can be used to
detect the phonon angular momentum of IR-active modes.

The paper is organized as follows. In Sec. \ref{sec:th} we develop the non-adiabatic linear response theory leading to the antisymmetric part of the dynamic infrared dielectric function. In Sec. \ref{sec:tec} we explain the technical details of the first principles time dependent density functional theory calculation. In sec. \ref{sec:results} we apply the theory and the methodological developments to the circular dichroic absorption and to the Faraday and Kerr effects in CrI$_3$. Finally we draw the conclusions.

\section{Theory\label{sec:th}}
\subsection{Definitions}
We consider a crystal composed of $N$ cells under periodic boundary
conditions, and with $N_a$ atoms in the unit cell. The position of an
atom is identified by the vector
\begin{equation}
\mathbf{R}_{L a} \equiv \mathbf{R}_L + {\boldsymbol \tau}_{a} + \mathbf{u}_{L a},
\end{equation}
where  $\mathbf{R}_L$ is a direct lattice vector, ${\boldsymbol \tau}_{a}$
the equilibrium position of the $a$-th atom in the
unit cell, and $\mathbf{u}_{La}$ indicates the deviation from equilibrium of
the nuclear position. The equilibrium positions are determined by the condition of vanishing forces, i.e. 
\begin{equation}
    {\bf F}_{L a}=-\left.\frac{\partial E(\mathbf{R})}{\partial \mathbf{R}_{L a} }\right|_{\mathbf{u}=\mathbf{0}}={\bf 0},
\end{equation} 
where $E(\mathbf{R})$ is the Born-Oppenheimer energy surface.


We introduce the adiabatic force-constant matrix
\begin{equation}
    C_{a\alpha,b\beta}(\mathbf{R}_L-\mathbf{R}_M)=\left.\frac{\partial^2 E(\mathbf{R})}{\partial R_{L a \alpha} \partial R_{Mb\beta}}\right|_{\mathbf{u}={\mathbf{0}}},
\end{equation}
where $\alpha$ and $\beta$ are Cartesian indices.

We define the Fourier transform of the atomic displacement from equilibrium as 
\begin{equation}
\mathbf{u}_{\mathbf{q}a}=\frac{1}{N}\sum_L \mathbf{u}_{L a} e^{-i\mathbf{q}\cdot\mathbf{R}_L},
\end{equation}
and similarly,
\begin{equation}
C_{a\alpha,b\beta}(\mathbf{q})=\frac{1}{N}\frac{\partial^2 E(\mathbf{R})}{\partial u_{\mathbf{q}a\alpha}^* \partial u_{\mathbf{q}b\beta}}.
\end{equation}
The  dynamical matrix in Fourier space is obtained by dividing the force constant matrix by the square root of the masses of the atoms. In the adiabatic approximation we find
\begin{equation}
D_{a\alpha,b\beta}(\mathbf{q})=\frac{C_{a\alpha,b\beta}(\mathbf{q})}{\sqrt{M_a}\sqrt{M_b}}.
\end{equation}
The diagonalization of the adiabatic dynamical matrix
in Fourier space leads to the adiabatic phonon frequencies
$\omega_{\mathbf{q}\nu}$ and eigenvectors
$\mathbf{e}_{\mathbf{q}\nu}$, namely
\begin{equation}
    \sum_{b\beta}\left[D_{a\alpha,b\beta}(\mathbf{q})-\omega^2_{\mathbf{q}\nu}\delta_{ab}\delta_{\alpha\beta}\right] e_{\mathbf{q}\nu}^{b\beta}=0
\end{equation}
where $e_{\mathbf{q}\nu}^{b\beta}=\sqrt{M_b}u_{\mathbf{q}\nu}^{b\beta}$
are the Cartesian components of the phonon eigenvectors. We assume the
eigenvectors to be normalized to unity. The adiabatic
phonon eigenvectors at the zone
center can be taken as  yielding
  linearly-polarized (nonchiral) phonons, and a symmetric (time-even) lattice dielectric function.\\

\subsection{Nonadiabatic dielectric function}

%
%
 An infrared  oscillating external electric field acts on an insulating
crystal of gap $\Delta$ at $t=0$.  More specifically, we
consider a monochromatic oscillating electric field of the form

\begin{equation}
{\mathcal E}_\alpha(t)={\mathcal E}_\alpha e^{-i\omega t} + \text{c.c.}
 \label{eq:field_mono}
 \end{equation}
where ``c.c.'' stands for complex conjugate and $\omega < \Delta$.  The
 electric field induces a monochromatic displacement in the ions. As a
 result,
\begin{eqnarray}
    \mathbf{u}_{La}(t)=\mathbf{u}_{La}e^{-i\omega t} + \text{c.c.}
    \label{eq:ions_mono}
\end{eqnarray}
In order to obtain the equations of motion of the ions under the action of the oscillating field,
we need to calculate the time-dependent retarded force constant
matrix, namely
\begin{eqnarray}
C_{a\alpha,b\beta}(\mathbf{q},t-t^\prime)=\frac{1}{N}\frac{\partial^2 E(\mathbf{R})}{\partial u_{\mathbf{q}a\alpha}^*(t^\prime) \partial u_{\mathbf{q}b\beta}(t)} \,\,\, {\rm } \theta(t-t^\prime)\nonumber \\
\end{eqnarray}
where $\theta(t-t^\prime)$ is the Heaviside function enforcing
 the retarded character of the response. 

The $\omega-$transform of the force constant matrix reads
\begin{eqnarray}
C_{a\alpha,b\beta}(\mathbf{q},\omega)=\int dt \,C_{a\alpha,b\beta}(\mathbf{q},t) \,e^{i\omega t}
\end{eqnarray}

The equation of motion for the Fourier transform of the ionic
displacement at the zone center ($\mathbf{q}=\mathbf{0}$)
 reads:
\begin{widetext}
\begin{equation}
    M_{a}\omega^2  u_{\mathbf{0}a\alpha}=\sum_{b\beta}C_{a\alpha,b\beta}(\omega) u_{\mathbf{0}b\beta}-i\hbar\omega
    \Gamma_a u_{\mathbf{0}a\alpha}
-\sum_{\beta}Z^*_{a\beta\alpha}{\mathcal E}_\beta.\label{eq:motion_omega}
\end{equation}
\end{widetext}
In
this expression $C_{a\alpha,b\beta}(\omega)$ denotes
$C_{a\alpha,b\beta}(\mathbf{0},\omega)$, and $Z^*_{a\beta\alpha}$ is
the Born effective charge tensor.
%
$\Gamma_a$ is the drag-force coefficient for atom $a$, and
$u_{\mathbf{0}a\alpha}$ is the Fourier transform of the coefficient
$u_{L a\alpha}$ on the right-hand side of Eq.~\eqref{eq:ions_mono}.

Within time dependent density functional theory and in the adiabatic
local density approximation, the 
  nonadiabatic force constant matrix for frequencies $\omega$ smaller
than the single-particle gap $\Delta$ has the form
\cite{CPMPhysRevB.82.165111}

\begin{equation}
C_{a\alpha,b\beta}(\omega)\approx C_{a\alpha,b\beta}(\omega=0)+\Pi_{a\alpha,b\beta}(\omega), \label{eq:CPM}
\end{equation}
with
\begin{widetext}
\begin{equation}
    \Pi_{a\alpha,b\beta}(\omega)=\frac{1}{N}
    \sum_{\mathbf{k}}\sum_{mn}\left[\frac{f_{\mathbf{k} m}-f_{\mathbf{k} n}}{E_{\mathbf{k}m}-E_{\mathbf{k}n}+\hbar\omega} -\frac{f_{\mathbf{k} m}-f_{\mathbf{k} n}}{E_{\mathbf{k}m}-E_{\mathbf{k}n}}
    \right]
    \left\langle u_{\mathbf{k}n}\left|\frac{\partial H_{\rm KS}}{\partial u_{\mathbf{0}a\alpha}}\right|u_{\mathbf{k}m}\right\rangle
\left\langle u_{\mathbf{k}m}\left|\frac{\partial H_{\rm KS}}{\partial u_{\mathbf{0}b\beta}}\right|u_{\mathbf{k}n}\right\rangle.
\label{eq:PI}
\end{equation}
\end{widetext}
Here $H_{\rm KS}$ is the static Kohn-Sham Hamiltonian in the absence
of the electric field, $E_{\mathbf{k}n}$ and $E_{\mathbf{k}m}$ are the
Kohn-Sham single-particle energies, $f_{\mathbf{k}n}$ is the Fermi
function, and $u_{\mathbf{k}n}$ is the cell-periodic part of the Bloch
function.

By substituting Eqs.~\eqref{eq:CPM} and~\eqref{eq:PI} into the
equations of motions and solving for the Cartesian components of the
$\mathbf{q}=\mathbf{0}$ phonon eigenvector
${ e}_{\mathbf{0}}^{a\alpha}$, we obtain
\begin{eqnarray}
   {e}^{a\alpha}_{\mathbf{0}} =\sum_{b\beta\gamma}
   \left[D(\omega)-\omega^2-i\omega {\tilde \Gamma}\right]^{-1}
   _{a\alpha,b\beta} \frac{ Z^{*}_{b\gamma\beta}{\mathcal E}_{\gamma}}{\sqrt{M}_b} \nonumber\\
   & &
\end{eqnarray}

where  we have defined ${\tilde \Gamma}_{a\alpha,b\gamma}=\frac{\Gamma_a}{M_a}\delta_{ab}\delta_{\alpha\gamma}$ and
\begin{equation}
D_{a\alpha,b\beta}(\omega)=\frac{C_{a\alpha,b\beta}(\omega)}{\sqrt{M_a}\sqrt{M_b}}.
\end{equation}

%

The $\alpha$ component of the electric polarization
induced by the optical electric field is given by:
\begin{equation}
P_\alpha=\sum_\beta \chi_{\alpha\beta}^{\rm el}{\mathcal E}_\beta+
\frac{1}{V_0}\sum_{a\beta}Z^*_{a\alpha\beta}
u_{\mathbf{0}a\beta}
\equiv\sum_\beta \chi_{\alpha\beta}{\mathcal E}_\beta,
\end{equation}
with $\chi_{\alpha\beta}^{\rm el}$ the electronic (clamped-ion)
susceptibility and $V_0$ the unit-cell volume.  It follows that the
dielectric tensor reads

\begin{widetext}
\begin{eqnarray}
\epsilon_{\alpha\beta}(\omega)&=&
\epsilon_{\alpha\beta}^{\infty}+
\frac{4\pi}{V_0}\sum_{ab}\sum_{\gamma\eta} \frac{Z_{a\alpha\gamma}^*}{\sqrt{M_a}}
\left[D(\omega)-\omega^2-i\omega{\tilde\Gamma}\right]^{-1}
_{a\gamma,b\eta}
\,\,\frac{Z_{b\beta\eta}^*}{\sqrt{M_b}},
\end{eqnarray}
\end{widetext}
where  $\epsilon_{\alpha\beta}^{\infty}=1+4\pi\chi_{\alpha\beta}^{\rm
    el}$.  We now define the adiabatic mode effective charge for the
phonon band $\nu$ at the zone center as
\begin{equation}
Z_{\nu\alpha}^*=\sum_{a\beta}Z_{a\alpha\beta}^* \frac{e^{a\beta}_{\mathbf{0}\nu}}{\sqrt{M_a}}.
\end{equation}
As the adiabatic zone-center phonon eigenvectors  $e^{a\beta}_{\mathbf{0}\nu}$ can be chosen as real,
$Z_{\nu\alpha}^*$ is also a real quantity.

We can then write the inner product in the dielectric function by
transforming it in the normal-mode basis,
\begin{eqnarray}
 \epsilon_{\alpha\beta}(\omega)&=&    \epsilon_{\alpha\beta}^{\infty}+\nonumber \frac{4\pi}{V_0}\sum_{\mu\nu} Z^*_{\mu\alpha}  
 \left[D(\omega)-\omega^2-i\omega\Gamma^0\right]^{-1}
 _{\mu\nu}
 Z^*_{\nu\beta}  \nonumber\\
 & &\label{eq:diel_tot}
\end{eqnarray}
where $\Gamma^0_{\mu\nu}=\sum_{a\alpha}\frac{e^{a\alpha}_{\mathbf{0}\mu} \Gamma_a e^{a\alpha}_{\mathbf{0}\nu}}{M_a} $. 

The  nonadiabatic dynamical
matrix can be expressed via Eq.~\eqref{eq:CPM} in the normal-mode basis
as
\begin{eqnarray}
D_{\mu\nu}(\omega)\approx D_{\mu\nu}(\omega=0)+\overline{\Pi}_{\mu\nu}(\omega),
\label{eq_CPM_for_D}
\end{eqnarray}\\
where $D_{\mu\nu}(\omega=0)$ is the adiabatic dynamical matrix
introduced earlier, which is diagonal on the basis of
the  adiabatic
eigenmodes, while the  nonadiabatic \cite{CPMPhysRevB.82.165111}
correction reads
\begin{widetext}
\begin{equation}
    \overline{\Pi}_{\mu\nu}(\omega)=\frac{1}{N}
    \sum_{\mathbf{k}}\sum_{n,m}\left[\frac{f_{\mathbf{k} m}-f_{\mathbf{k} n}}{E_{\mathbf{k}m}-E_{\mathbf{k}n}+\hbar\omega} -\frac{f_{\mathbf{k} m}-f_{\mathbf{k} n}}{E_{\mathbf{k}m}-E_{\mathbf{k}n}}
    \right]
    \left\langle u_{\mathbf{k}n}\left|\frac{\partial H_{\rm KS}}{\partial \mathbf{e}_{\mathbf{0}\mu}}\right|u_{\mathbf{k}m}\right\rangle
\left\langle u_{\mathbf{k}m}\left|\frac{\partial H_{\rm KS}}{\partial
\mathbf{e}_{\mathbf{0}\nu}}
\right|u_{\mathbf{k}n}\right\rangle.
\label{eq:def_Pi_bar}
\end{equation}
\end{widetext}
It is worthwhile underlying that Eq. \ref{eq:diel_tot} is the
 nonadiabatic version of Eq.~(50) in
Ref. \cite{Gonze_LeePhysRevB.55.10355}.

The antisymmetric  part of the dielectric
function is obtained  from
Eq.~\eqref{eq:epsil_AS_def}.
%

%
As $D_{\mu\nu}(0)$ , $\Gamma^0_{\mu\nu}$ and
$\epsilon_{\alpha\beta}^{\infty}$ are real quantities, by substituting
Eqs. \eqref{eq:diel_tot}, \eqref{eq_CPM_for_D} and
\eqref{eq:def_Pi_bar} into Eq.~\eqref{eq:epsil_AS_def} we obtain

\begin{widetext}
\begin{equation}
\epsilon_{\alpha\beta}^{\rm AS} (\omega)= 
\epsilon_{\alpha\beta}^{\rm AS,\infty}
+\frac{2\pi}{V_0}\sum_{\mu\nu} Z_{\mu\alpha}^* \left\{
\left[
D(0)+\overline{\Pi}(\omega)-\omega^2 -i\omega\Gamma^0
\right]^{-1}-
\left[
D(0)+\overline{\Pi}^*(\omega)-\omega^2 -i\omega\Gamma^0
\right]^{-1}
\right\}_{\mu\nu} Z_{\nu\beta}^*.
\end{equation}
\end{widetext}
For $\omega \ll \Delta$, the electronic dielectric tensor
$\epsilon^\infty_{\alpha\beta}$ is Hermitean (non-dissipative); thus
its real part is symmetric and its imaginary part is
antisymmetric. The reality condition
  $\left[\epsilon^\infty_{\alpha\beta}(\omega)\right]^*=\epsilon^\infty_{\alpha\beta}(-\omega)$
  implies that the imaginary-antisymmetric part is odd in frequency,
and therefore it vanishes for $\omega \ll \Delta$. Hence,
${\rm Im}\,\epsilon^{{\rm AS},\infty}_{\alpha\beta}$ is negligible in
the infrared regime, and can be safely dropped. In the adiabatic approximation,
as $\overline{\Pi}(0)=0$, it follows that
$\epsilon_{\alpha\beta}^{\rm AS} =0$ when the system is
insulating.
%
%
We complete this theoretical part by noting that the real (dissipative
dichroic) and imaginary (reactive dichroic) parts are, respectively,
given by:
\begin{widetext}
\begin{eqnarray}
\Re\epsilon_{\alpha\beta}^{\rm AS} (\omega)&=& 
\Re\epsilon_{\alpha\beta}^{\rm AS,\infty}
+\frac{\pi}{V_0}\sum_{\mu\nu} Z_{\alpha\mu}^* \left\{
\left[
D(0)+\overline{\Pi}(\omega)-\omega^2 -i\omega\Gamma^0
\right]^{-1}-
\left[
D(0)+\overline{\Pi}^*(\omega)-\omega^2 -i\omega\Gamma^0
\right]^{-1}
\right\}_{\mu\nu} Z_{\beta\nu}^*\nonumber \\
& &+\frac{\pi}{V_0}\sum_{\mu\nu} Z_{\alpha\mu}^* \left\{
\left[
D(0)+\overline{\Pi}^*(\omega)-\omega^2 +i\omega\Gamma^0
\right]^{-1}-
\left[
D(0)+\overline{\Pi}(\omega)-\omega^2 +i\omega\Gamma^0
\right]^{-1}
\right\}_{\mu\nu} Z_{\beta\nu}^*\\
\Im\epsilon_{\alpha\beta}^{\rm AS} (\omega)&=& 
\frac{\pi}{V_0}\sum_{\mu\nu} Z_{\alpha\mu}^* \left\{
\left[
D(0)+\overline{\Pi}(\omega)-\omega^2 -i\omega\Gamma^0
\right]^{-1}-
\left[
D(0)+\overline{\Pi}^*(\omega)-\omega^2 -i\omega\Gamma^0
\right]^{-1}
\right\}_{\mu\nu} Z_{\beta\nu}^*\nonumber \\
& &+\frac{\pi}{V_0}\sum_{\mu\nu} Z_{\alpha\mu}^* \left\{
\left[
D(0)+\overline{\Pi}^*(\omega)-\omega^2 +i\omega\Gamma^0
\right]^{-1}-
\left[
D(0)+\overline{\Pi}(\omega)-\omega^2 +i\omega\Gamma^0
\right]^{-1}
\right\}_{\mu\nu} Z_{\beta\nu}^*,
\end{eqnarray}
\end{widetext}
where again $\omega \ll \Delta$ is assumed.

\section{Technical details\label{sec:tec}}

\begin{figure}[h]
\includegraphics[width=\columnwidth]{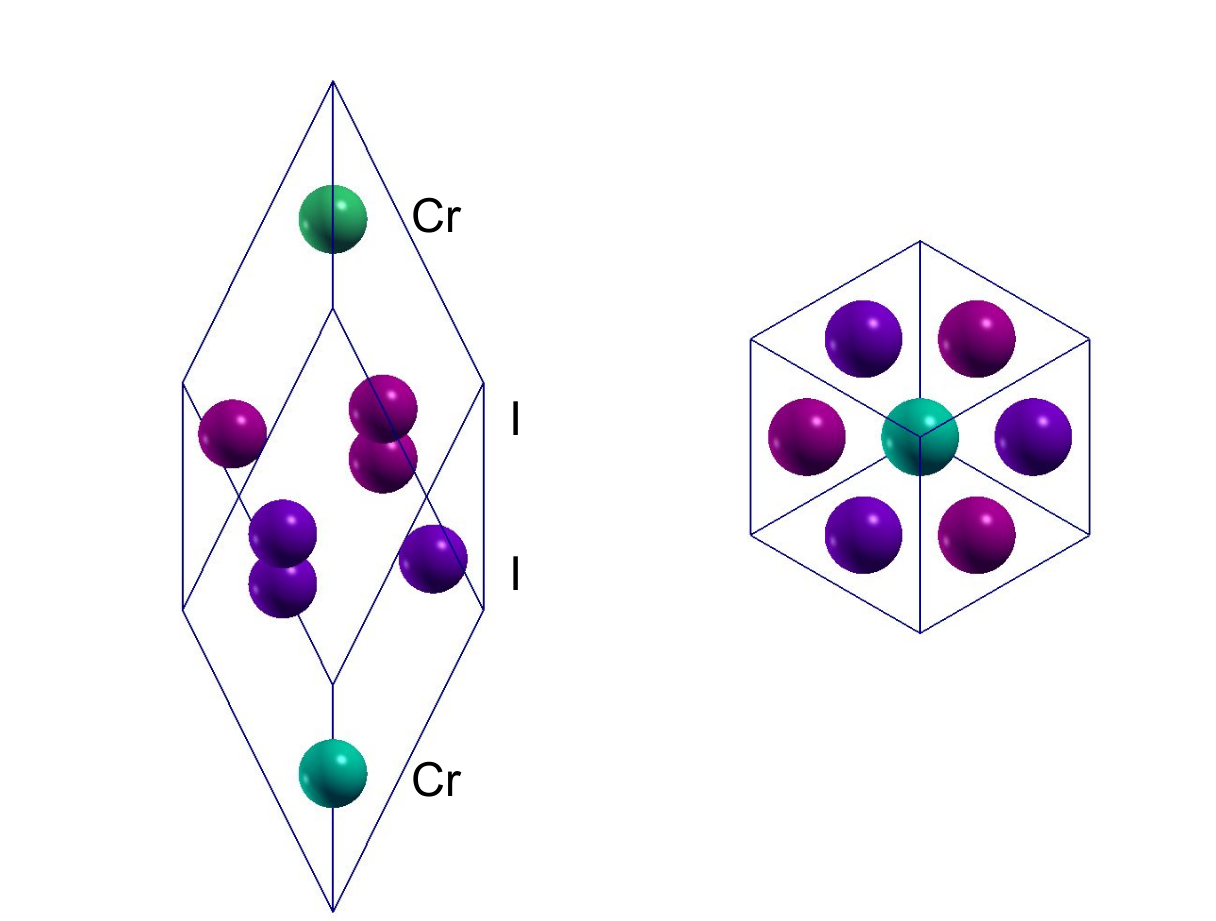}
    \caption{Crystal structure of CrI$_3$.}
    \label{fig:struct}
\end{figure}

We simulate the structural, electronic and vibrational properties of CrI$_3$ by using the \textsc{quantum ESPRESSO} (QE) code \cite{QE-2017}. We adopt the PBE exchange and correlation functional \cite{PBEPhysRevLett.78.1396}. We use ONCV \cite{ONCVPhysRevB.88.085117} fully relativistic norm-conserving pseudopotentials for Cr and I from the PseudoDojo \cite{VANSETTEN201839} repository, option accuracy stringent.  CrI$_3$ crystallizes in the space group $R\bar{3}$ (see Fig. \ref{fig:struct}). We use the experimental lattice parameters $\approx7.70 $\AA and the rhombohedral angle $\cos(\theta)\approx 0.6025$  ($\theta\approx 52.93^o$) and we perform structural optimization of internal coordinates. In the geometrical optimization of the internal coordinates we use a $4\times 4\times 4$ electron momentum mesh and a Fermi-Dirac smearing of $3\times 10^{-4}$ Ryd. We use a $120$ Ry cutoff on the kinetic energy and a $4$ times larger cutoff for the charge density.


The adiabatic phonon frequencies and the Born effective charge tensor
were calculated by using density functional perturbation theory (DFPT)
\cite{BaroniRevModPhys.73.515}.  For technical reasons, the dielectric
tensor $\epsilon_{\alpha\beta}^{\infty}$ and Born effective charge
tensor $Z_{a\alpha\beta}^*$ were calculated in the collinear spin
approximation, while the adiabatic phonon frequencies were obtained
via a fully relativistic noncollinear DFPT calculation. We do not
expect substantial qualitative differences with respect to the case in
which the Born effective charges are calculated in the noncollinear
fully relativistic case.  The calculated
$\epsilon_{\alpha\beta}^{\infty}$ tensors are reported in
Tabs. \ref{tab:epsilon_infinity}.

In this work we focus on the infrared active E$_u$ mode at $216.4$
cm$^{-1}$, in the static harmonic approximation. This mode is twofold
degenerate in the adiabatic relativistic noncollinear magnetic case.

The nonadiabatic dynamical matrix $D_{a\alpha,b\beta}(\omega)$
was obtained via Eq. \ref{eq_CPM_for_D} using the method developed in Ref. \cite{BistoniPhysRevLett.126.225703} and in the fully relativistic approach by using a $8\times 8 \times 8$ electron-momentum grid.
The nonadiabatic effects redshift and induce a  $0.22$ cm$^{-1}$ splitting between the two $E_u$ modes, which frequencies become $214.70$ cm$^{-1}$ and $214.92$, respectively.
\begin{table}[h]
    \centering
    \begin{tabular}{|l|c|c|c|}
    \hline
$\epsilon_{\alpha\beta}^{\infty}$  &  $\beta=x$ & $\beta=y$  & $\beta=z$  \\\hline 
$\alpha=x$ & 7.97 & 0.0 & 0.0 \\
$\alpha=y$ & 0.0 & 7.97 & 0.0 \\
$\alpha=z$ & 0.0 & 0.0 & 6.07 \\ \hline
    \end{tabular}
    \caption{Dielectric tensor $\epsilon_{\alpha\beta}^{\infty}$
      calculated in density functional perturbation theory in the
      collinear spin approximation.}
    \label{tab:epsilon_infinity}
\end{table}

\section{Results\label{sec:results}}

\subsection{Dielectric function of CrI$_3$}
\begin{figure}[t]
\includegraphics[width=\columnwidth]{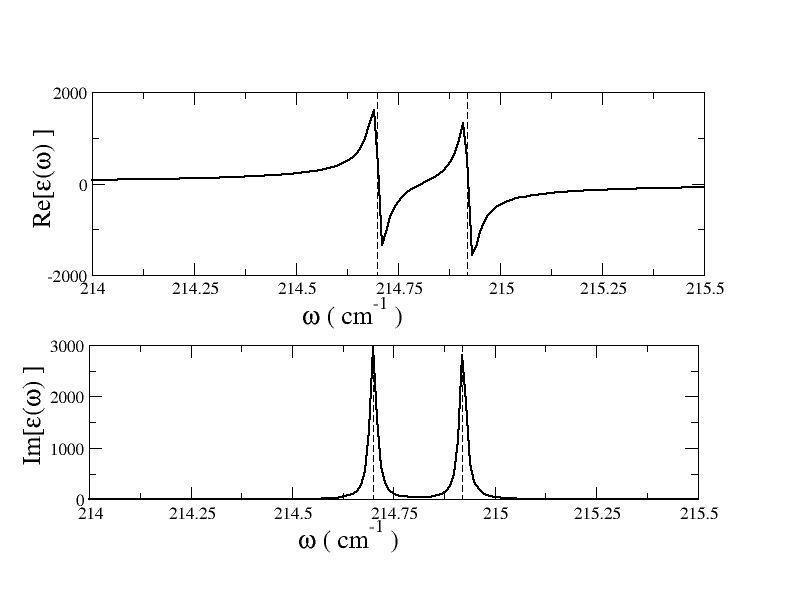}
\caption{Real and imaginary part of
  $\epsilon(\omega)=\frac{1}{2}[\epsilon_{xx}(\omega)+\epsilon_{yy}(\omega)]$. The
  curves have been broadened with a width of 0.02 cm$^{-1}$. The
  vertical dashed lines label the phonon frequencies split by the
  dynamical effects.}
    \label{fig:epsilon_w}
\end{figure}

The calculated real and imaginary parts of the average of the in-plane components of the dielectric function, i.e. $\epsilon(\omega)=\frac{1}{2}[\epsilon_{xx}(\omega)+\epsilon_{yy}(\omega)]$,
are shown in Fig. \ref{fig:epsilon_w}. The imaginary part peaks at the phonon frequencies splitted by dynamical effects (vertical dashed lines).
 The splitting between the phonon frequencies induced by the occurrence of phonon chirality is  $\approx 0.22$ cm$^{-1}$. Nevertheless, the phonon angular momentum associated with each mode is not small, as we evaluate it to be $\approx\pm \hbar/2$, in agreement with what reported in \cite{RenPhysRevLett.130.086701}. Thus, even for large phonon angular momenta, the induced splitting can be quite small, as already shown in Ref. \cite{BistoniPhysRevLett.126.225703}.

The splitting measured in experiments, however, strongly depends on the energy difference occurring between the infrared phonon frequencies and the magnon excitation. In our formalism, as we have neglected the $\omega-$dependence of the electron-phonon vertex (we use TDDFT with a static electron-phonon vertex), the splitting is the one at frozen magnetic moments, which is necessarily different from the experimental one.

\begin{figure}
\includegraphics[width=\columnwidth]{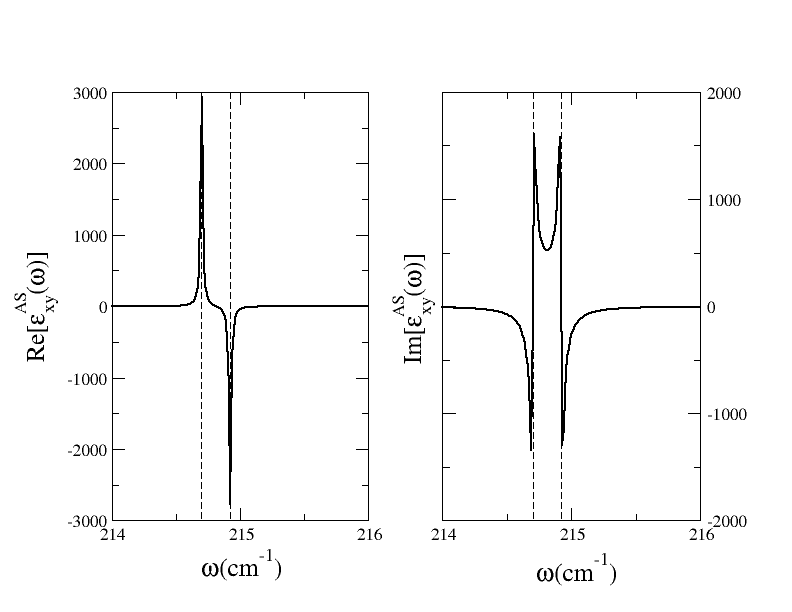}
    \caption{Real and imaginary part of $\epsilon_{xy}^{AS}(\omega)=(\epsilon_{xy}(\omega)-\epsilon_{yx}(\omega))/2$. The curves have been broadened with a width of 0.02 cm$^{-1}$. The dashed lines mark the phonon frequencies splitted by dynamical effects. }    \label{fig:epsilon_xy_AS_w}
\end{figure}

The dissipative (real part) and reactive (imaginary part) dichroic of
the dielectric tensor $\epsilon_{xy}^{AS}(\omega)$ are shown in
Fig. \ref{fig:epsilon_xy_AS_w}. Both parts display marked resonances
at the position of the two splitted infrared phonon frequencies. We
recall that in the absence of chirality or dynamical effects, i.e.
$\overline{\Pi}(\omega)=0$, bot the real and imaginary parts of
$\epsilon_{xy}^{\rm AS}(\omega)$ would be zero.

\subsection{Refractive index}

We consider light impinging on a sample. We assume the fields to be polarized within the $xy$ plane and propagating along the $z$ direction, ${\bf E}e^{i(kz-\omega t)}$ and ${\bf H}e^{i(kz-\omega t)}$.
We define the refractive index as $n=ck/\omega$. The fields inside the material satisfy the following relations:
\begin{equation}
\label{eq:H_n_E}
{\bf H}=n\mathbf{\hat{z}}\times{\bf E}
\end{equation}
and
\begin{equation}
\label{eq:D_n_H}
\mathbf{D}=\bm{\varepsilon} {\bf E}=-n\mathbf{\hat{z}}\times{\bf H}\,.
\end{equation}
In vacuum, $n=1$.
if we combine ${\bf D}=-{\bf n}\times{\bf H}$ with $\mathbf{n}=n\mathbf{\hat{z}}$ and ${\bf H}={\bf n}\times{\bf E}$, we arrive at
\begin{align}
    \label{eq:D_LL}
    {\bf D}=n^2\, {\bf E}-({\bf n}\cdot{\bf E})\,{\bf n}\,,
\end{align}
which can be written as
\begin{eqnarray}
    \label{eq:D_LL2}
    D_x&=&n^2\, E_x=n^2 \left({\varepsilon}^{-1}_{xx} D_x+{\varepsilon}^{-1}_{xy} D_y\right)\nonumber \\
    D_y&=&n^2\, E_y=n^2 \left({\varepsilon}^{-1}_{yx} D_x+{\varepsilon}^{-1}_{yy} D_y\right)\nonumber\\
    D_z&=&0\,.
\end{eqnarray}
or, in matrix form,
\begin{equation}
    \begin{pmatrix}
        \varepsilon^{-1}_{xx}-n_a^{-2} & \varepsilon^{-1}_{xy} \\
        \varepsilon^{-1}_{yx} & \varepsilon^{-1}_{yy}-n_a^{-2}
    \end{pmatrix}
    \begin{pmatrix}
        D_{a,x} \\
        D_{a,y}
    \end{pmatrix}
    =
    \begin{pmatrix}
        0\\
        0
    \end{pmatrix}
    \,.
\label{eq:epsilon_inverse_linear_system}
\end{equation}
which is an eigenvalue problem of a complex and non-Hermitian $2\times2$ matrix. We label the eigenvectors and the corresponding eigenvalues as ${\bf D}_a=\bm{\varepsilon} {\bf E}_a$ and $n_a^{-2}$, respectively ($a=1,\,2$). Both eigenvalues and eigenvectors are  complex.

Therefore,
\begin{equation}
n_a^{-2}=\frac{\varepsilon^{-1}_{xx}+\varepsilon^{-1}_{yy}}{2}\pm\sqrt{\left(\frac{\varepsilon^{-1}_{xx}-\varepsilon^{-1}_{yy}}{2}\right)^2+\varepsilon^{-1}_{xy}\,\varepsilon^{-1}_{yx}}\,,
\label{eq:n_a_2}
\end{equation}

There are two possible values for each refractive index $n_a(\omega)$, as $n_a(\omega)=\pm 1/\sqrt{n_a^{-2}(\omega)}$. Assuming that the vacuum is in $z<0$ and the material is in $z>0$, $n_a$ must satisfy both ${\rm Re}\,n_a>0$ (propagating along the $+z$ direction in the material) and ${\rm Im}\,n_a>0$ (decaying in the material). 

The eigenvalue $n_a^{-2}$  satisfies the causality relation, that is, the imaginary part of $n_a^{-2}$ is negative to make the imaginary part of $n_a^2$ positive. In order to demonstrate this point, we assume that the $3\times3$ dielectric function  is written as a sum of two terms, $\bm{\varepsilon}=\bm{\varepsilon}_1+i\,\bm{\varepsilon}_2$, where ${\bm\varepsilon}_1$ and ${\bm\varepsilon}_2$ are both hermitian. Then, for arbitrary ${\bf E}$ and ${\bf D}$ vectors satisfying ${\bf D}=\bm{\varepsilon}{\bf E}$, we have  ${\bf D}^\dagger \bm{\varepsilon}^{-1}{\bf D}={\bf E}^\dagger \bm{\varepsilon}^{\dagger}{\bf E}={\bf E}^\dagger (\bm{\varepsilon}_1-i\bm{\varepsilon}_2){\bf E}$. Because $\bm{\varepsilon}_2$ is positive-definite,
\begin{equation}
{\rm Im}\,{\bf D}^\dagger \bm{\varepsilon}^{-1}{\bf D}={\bf E}^\dagger\left(-\bm{\varepsilon}_2\right){\bf E}<0
    \label{eq:ImD_epsinv_D}
\end{equation}
always. Now, if we choose ${\bf D}$ to be an eigenvector satisfying Eqs.~\eqref{eq:D_LL2} and~\eqref{eq:epsilon_inverse_linear_system},
\begin{equation}
{\bf D}_a^\dagger \bm{\varepsilon}^{-1}{\bf D}_a=n_a^{-2}\,{\bf D}_a^\dagger{\bf D}_a\,.
    \label{eq:Da_epsinv_Da}
\end{equation}
Plugging Eq.~\eqref{eq:Da_epsinv_Da} into Eq.~\eqref{eq:ImD_epsinv_D}, we arrive at ${\rm Im}\,n_a^{-2}<0$, completing the proof.
This condition is equivalent to stating that the imaginary part of the eigenvalues of $\bm{\varepsilon}$ is positive.

\begin{figure}
\includegraphics[width=\columnwidth]{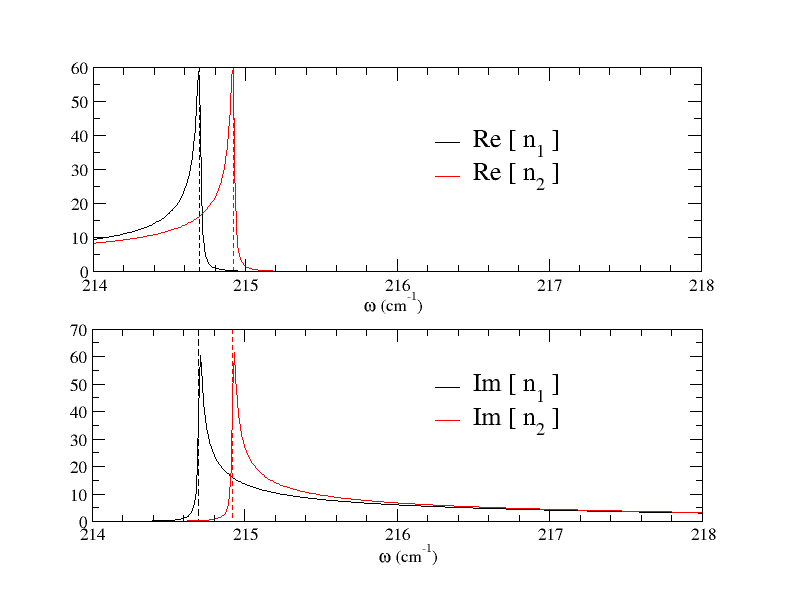}
\caption{Real and imaginary parts of the refractive index
  $n(\omega)$. The vertical dashed lines marks the energy of the
  infrared phonons split by dynamical
  effects. }  \label{fig:refraction_index}
\end{figure}

The calculated real and imaginary parts of the dynamical refraction index $n_a(\omega)$ are shown in Fig. \ref{fig:refraction_index}. In Fig. \ref{fig:refraction_index} we choose the solution of the equation $n_a(\omega)=\pm1/\sqrt{n_a^{-2}(\omega)}$ respecting the dissipative character of $n_a(\omega)$ inside the material (i.e. with 
$\Im n_a(\omega) > 0$).
As it can be seen, left and right polarized light experience two different refraction indexes that will generate a nonzero circular dichroic absorption as well as nonzero Faraday and Kerr rotations.

\subsection{Reflection from an infinitely thick sample}

In this section we  obtain the equations for the reflectivity and the circular dichroic absorption from a measurement in reflection from a sample assumed to have an infinite thickness.

We define the left and right polarization as $\mathbf{\hat{e}}_{\pm 1}=\frac{\mathbf{\hat{x}}\pm i\mathbf{\hat{y}}}{\sqrt{2}}$, where 
$\mathbf{\hat{x}}$ and $\mathbf{\hat{y}}$ are the orthogonal unit vectors in the $xy$-plane.
We define the incident, reflected, and transmitted waves for circularly polarized incidence (with handedness $s=\pm1$) as follows:
\begin{align}
\label{eq:E_H_i_r_t_circular}
{\bf E}_{i,s} =&\mathbf{\hat{e}}_s\,e^{i\omega z/c}\,e^{-i\omega t}
\nonumber \\
{\bf E}_{r,s} =& \left(\mathbf{\hat{e}}_{+1}\,r_{+1,s}+\mathbf{\hat{e}}_{-1}\,r_{-1,s}\right) \,e^{-i\omega z/c}\,e^{-i\omega t}
\nonumber \\
{\bf E}_{t,s} =& {\bf E}_1\,t_{1,s} \,e^{i\omega n_1\,z/c}+{\bf E}_2\,t_{2,s} \,e^{i\omega n_2\,z/c}\,e^{-i\omega t}
ì\,.
\end{align}
The incident, reflected and transmitted magnetic fields can be obtained from 
${\bf H}=-i\frac{c}{\omega}\nabla\times{\bf E}$.

The coefficients $r_{+1,s}$, $r_{-1,s}$, $t_{1,s}$, and $t_{2,s}$ are determined from the conditions that the tangential ($xy$) components of  the fields ${\bf E}$ and ${\bf H}$ are continuous across the boundary, namely:
\begin{align}
\label{eq:bcs_circular}
\mathbf{\hat{e}}_s+\mathbf{\hat{e}}_{+1}\,r_{+1,s}+\mathbf{\hat{e}}_{-1}\,r_{-1,s}&={\bf E}_1\,t_{1,s}+{\bf E}_2\,t_{2,s}
\nonumber \\
\mathbf{\hat{e}}_s-\mathbf{\hat{e}}_{+1}\,r_{+1,s}-\mathbf{\hat{e}}_{-1}\,r_{-1,s}&=n_1\,{\bf E}_1\,t_{1,s}+n_2\,{\bf E}_2\,t_{2,s}\,,
\end{align}
From the solution of this $8$ dimensional linear system we determine all the necessary coefficients. We underline that  $t_{1,s}$ and $t_{2,s}$ do depend on the magnitudes of ${\bf E}_1$ and ${\bf E}_2$, respectively; however, their products $t_{1,s}\,{\bf E}_{1}$ and $t_{2,s}\,{\bf E}_{2}$ do not. 

\begin{figure}
\includegraphics[width=\columnwidth]{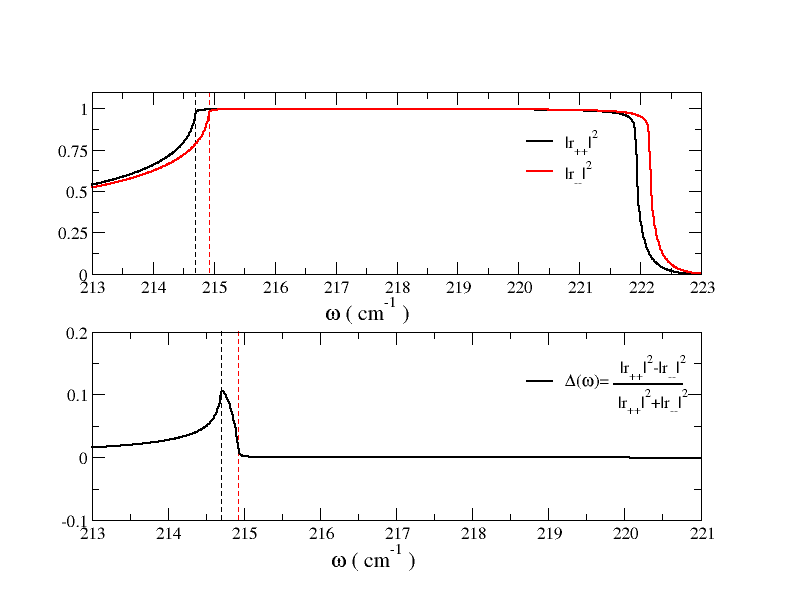}
\caption{Reflectivity (top) and Dichroic absorption (bottom) in
  CrI$_3$ in the assumption of a measurement in reflection from an
  infinite sample. The vertical dashed lines marks the energy of the
  infrared phonons split by dynamical
  effects. }  \label{fig:dichroic_reflection}
\end{figure}

The reflectivities for two different circular polarizations are $\left|r_{+1,+1}\right|^2$ and $\left|r_{-1,-1}\right|^2$ and they are plotted in the top panel of Fig. \ref{fig:dichroic_reflection}. The calculated reflectivity  agrees qualitatively and semiquantitatively with experiments \cite{ma16144909,Tomarchio2021}.

The dichroic absorption signal is then
\begin{equation}
\Delta(\omega)=\frac{\left|r_{+1,+1}\right|^2-\left|r_{-1,-1}\right|^2}{\left|r_{+1,+1}\right|^2+\left|r_{-1,-1}\right|^2}
\label{eq:dichroic_reflection}
\end{equation}

The result of Eq. \ref{eq:dichroic_reflection} 
is shown in Fig. \ref{fig:dichroic_reflection}. As it can be seen, even if the splitting is very small as the adiabatic local density approximation  employed in the calculation, the circular dichroic absorption is sizable and can be measured.

\subsection{Kerr rotation from the reflection of an infinite thick sample}

Kerr rotation is the rotation of the polarization of light when light is reflected from a magnetic material or in the presence of  an external magnetic field.
In order to model this effect,
we assume that the incident light has polarization $\mathbf{\hat{e}}_\theta=\mathbf{\hat{x}}\,\cos\theta+\mathbf{\hat{y}}\,\sin\theta$. Then, electric fields can be written as
\begin{align}
\label{eq:E_H_i_r_t_linear}
{\bf E}_{i,\theta} =&\mathbf{\hat{e}}_\theta\,e^{i\omega z/c}e^{-i\omega t}
\nonumber \\
{\bf E}_{r,\theta} =& \left(\mathbf{\hat{x}}\,r_{x,\theta}+\mathbf{\hat{y}}\,r_{y,\theta}\right) \,e^{-i\omega z/c}e^{-i\omega t}
\nonumber \\
{\bf E}_{t,\theta} =& {\bf E}_1\,t_{1,\theta} \,e^{i\omega n_1\,z/c}+{\bf E}_2\,t_{2,\theta} \,e^{i\omega n_2\,z/c}e^{-i\omega t}
\,.
\end{align}
Using the same boundary conditions as in the case of the circular polarization, we arrive at
\begin{align}
\label{eq:bcs_linear}
\mathbf{\hat{e}}_\theta+\mathbf{\hat{x}}\,r_{x,\theta}+\mathbf{\hat{y}}\,r_{y,\theta}&={\bf E}_1\,t_{1,\theta}+{\bf E}_2\,t_{2,\theta}
\nonumber \\
\mathbf{\hat{e}}_\theta-\mathbf{\hat{x}}\,r_{x,\theta}-\mathbf{\hat{e}}\,r_{y,\theta}&=n_1\,{\bf E}_1\,t_{1,\theta}+n_2\,{\bf E}_2\,t_{2,\theta}\,,
\end{align}
The solution of this $4-$dimensional linear system
leads to $r_{x,\theta}$, $r_{y,\theta}$, $t_{1,\theta}$, and $t_{2,\theta}$.

The angle of the polarization and the ellipticity can be obtained directly from these parameters as follows. We define the three Stokes parameters $s_1$, $s_2$ and $s_3$ as
\begin{align}
\label{eq:Stokes_parameters}
&s_1=I_{0^\circ}-I_{90^\circ}=\left|r_{x,\theta}\right|^2-\left|r_{y,\theta}\right|^2
\nonumber \\
&s_2=I_{+45^\circ}-I_{-45^\circ}=\left|\frac{r_{x,\theta}+r_{y,\theta}}{\sqrt{2}}\right|^2-\left|\frac{r_{x,\theta}-r_{y,\theta}}{\sqrt{2}}\right|^2
\nonumber \\
&s_3=I_{\rm RCP}-I_{\rm LCP}=\left|\frac{r_{x,\theta}+i\,r_{y,\theta}}{\sqrt{2}}\right|^2-\left|\frac{r_{x,\theta}-i\,r_{y,\theta}}{\sqrt{2}}\right|^2\,,
\end{align}
where we have used the polarizations for RCP and LCP are $\mathbf{\hat{e}}_{\rm RCP}=\mathbf{\hat{e}}_{-1}=\frac{\mathbf{\hat{x}}-i\mathbf{y}}{\sqrt{2}}$ and $\mathbf{\hat{e}}_{\rm LCP}=\mathbf{\hat{e}}_{+1}=\frac{\mathbf{\hat{x}}+i\,\mathbf{\hat{y}}}{\sqrt{2}}$, respectively.

Then, the polarization angle is given by
\begin{equation}
    \label{eq:pol_direction}
    \psi=\frac{1}{2}\tan^{-1}\frac{s_2}{s_1}\,.
\end{equation}
and the ellipticity can be obtained from
\begin{equation}
    \label{eq:ellipticity}
    \frac{b}{a}=\tan\left(\frac{1}{2}\tan^{-1}\frac{s_3}{\sqrt{s_1^2+s_2^2}}\right)\,,
\end{equation}
where $a$ and $b$ are the major and minor axes of an ellipse.
\begin{figure}
\includegraphics[width=1.1\columnwidth]{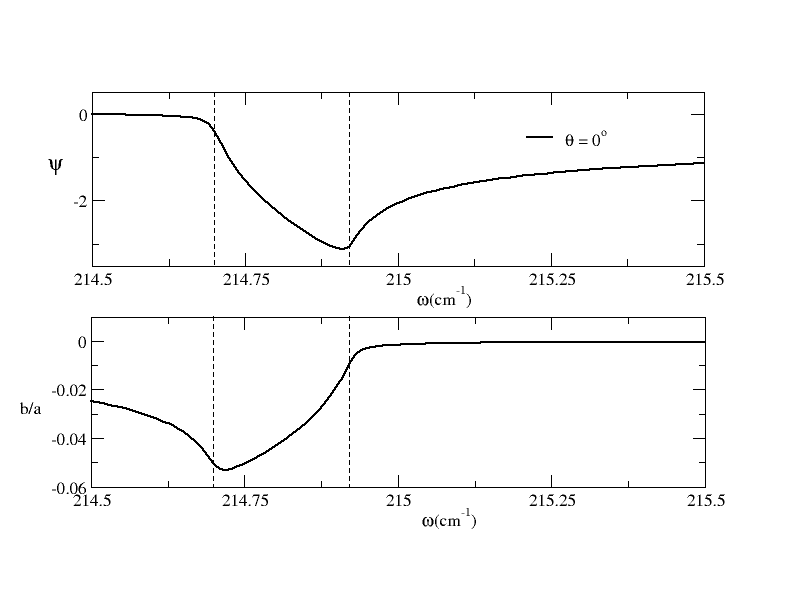}
\caption{Polarization angle (top) and ellipticity (bottom) for Kerr
  rotation as obtained from the reflection from an infinite thick
  sample of CrI$_3$ for $\theta=0^o$. The vertical dashed lines marks
  the energy of the infrared phonons split by dynamical
  effects. }  \label{fig:Faraday_reflection}
\end{figure}
The solution of the linear system is shown in Fig. \ref{fig:Faraday_reflection} for $\theta=0^o$. 

\subsection{Dichroism of light transmitted through a slab of thickness $d$}

As done in the previous sections, we write down the fields inside a sample of finite thickness $d$ in a point $0<z<d$. We neglect multiple reflection and consider $4$ electric fields, namely the incident ${\bf E}_{i,s}$, the reflected ${\bf E}_{r,s}$, the transmitted ${\bf E}_{t,s}$ and the field inside the medium ${\bf E}_{m,s} $. For 
the incident, reflected, intermediate ($0<z<d$), and transmitted fields we have :
\begin{align}
\label{eq:E_H_i_r_m_t_circular}
{\bf E}_{i,s} =&\mathbf{\hat{e}}_s\,e^{i\omega z/c}e^{-i\omega t}
\nonumber \\
{\bf E}_{r,s} =& \left(\mathbf{\hat{e}}_{+1}\,r_{+1,s}+\mathbf{\hat{e}}_{-1}\,r_{-1,s}\right) \,e^{-i\omega z/c}e^{-i\omega t}
\nonumber \\
{\bf E}_{m,s} =& {\bf E}_1\left(m_{11,s} \,e^{i\omega n_1\,z/c}+m_{12,s} \,e^{-i\omega n_1\,z/c}\right)e^{-i\omega t}
\nonumber \\
&+ {\bf E}_2\left(m_{21,s} \,e^{i\omega n_2\,z/c}+m_{22,s} \,e^{-i\omega n_2\,z/c}\right)e^{-i\omega t}
\nonumber \\
{\bf E}_{t,s} =& \left(\mathbf{\hat{e}}_{+1}\,t_{+1,s}+\mathbf{\hat{e}}_{-1}\,t_{-1,s}\right) \,e^{i\omega z/c}e^{-i\omega t}
\,.
\end{align}
The boundary conditions at $z=0$ and $z=d$ give
\begin{align}
\label{eq:bcs_circular2}
&\mathbf{\hat{e}}_s+\mathbf{\hat{e}}_{+1}\,r_{+1,s}+\mathbf{\hat{e}}_{-1}\,r_{-1,s}
\nonumber \\
&={\bf E}_1\,(m_{11,s}+m_{12,s})+{\bf E}_2\,(m_{21,s}+m_{22,s})\,,
\nonumber \\
&\mathbf{\hat{e}}_s-\mathbf{\hat{e}}_{+1}\,r_{+1,s}-\mathbf{\hat{e}}_{-1}\,r_{-1,s}
\nonumber \\
&=n_1\,{\bf E}_1\,(m_{11,s}-m_{12,s})+n_2\,{\bf E}_2\,(m_{21,s}-m_{22,s})
\end{align}
and
\begin{align}
\label{eq:bcs_circular3}
& {\bf E}_1\left(m_{11,s} \,e^{i\omega n_1\,d/c}+m_{12,s} \,e^{-i\omega n_1\,d/c}\right)
\nonumber \\
&+ {\bf E}_2\left(m_{21,s} \,e^{i\omega n_2\,d/c}+m_{22,s} \,e^{-i\omega n_2\,d/c}\right)
\nonumber \\
&= \left(\mathbf{\hat{e}}_{+1}\,t_{+1,s}+\mathbf{\hat{e}}_{-1}\,t_{-1,s}\right) \,e^{i\omega d/c}\,,
\nonumber \\
& n_1\,{\bf E}_1\left(m_{11,s} \,e^{i\omega n_1\,d/c}-m_{12,s} \,e^{-i\omega n_1\,d/c}\right)
 \nonumber \\
&+ n_2\,{\bf E}_2\left(m_{21,s} \,e^{i\omega n_2\,d/c}-m_{22,s} \,e^{-i\omega n_2\,d/c}\right)
\nonumber \\
&= \left(\mathbf{\hat{e}}_{+1}\,t_{+1,s}+\mathbf{\hat{e}}_{-1}\,t_{-1,s}\right) \,e^{i\omega d/c}\,,
\end{align}
respectively. We now have eight unknowns, $r_{+1,s}$, $r_{-1,s}$, $m_{11,s}$, $m_{12,s}$, $m_{21,s}$, $m_{22,s}$, $t_{+1,s}$, and $t_{-1,s}$, and eight equations [Eqs.~\eqref{eq:bcs_circular2} and~\eqref{eq:bcs_circular3}].
\begin{figure}
\includegraphics[width=1.1\columnwidth]{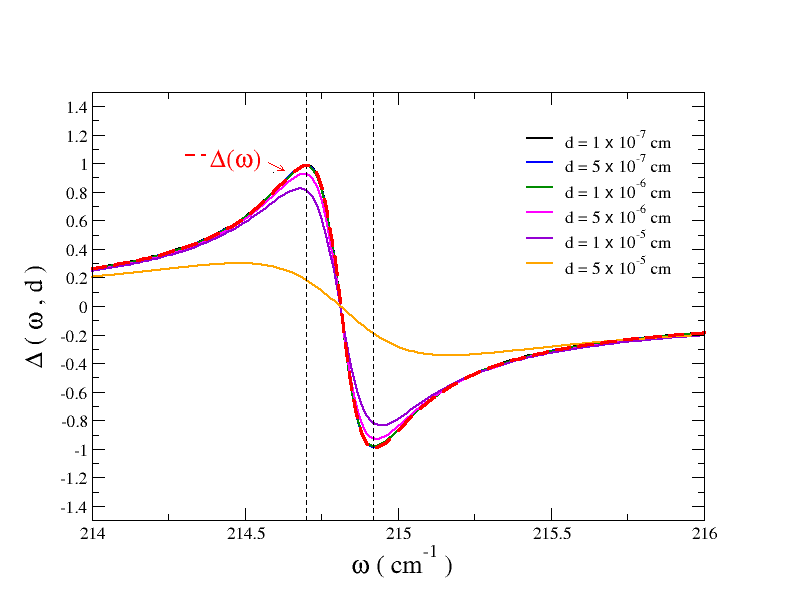}
    \caption{Dichroic absorption in  CrI$_3$ in the assumption of a measurement in transmission from a sample of thickness $d$. The red continous line is obtained from Eq. \ref{eq:Delta2}.
    The vertical dashed lines mark the energy of the infrared phonons splitted by dynamical effects.  }    \label{fig:Delta_tr}
\end{figure}
The circular dichroism for absorbance (neglecting internal reflections) is obtained as:
\begin{equation}
\label{eq:Delta}
\Delta(\omega,\,d)=\frac{\left(1-|t_{+1,+1}|^2\right)-\left(1-|t_{-1,-1}|^2\right)}{\left(1-|t_{+1,+1}|^2\right)+\left(1-|t_{-1,-1}|^2\right)}\,,
\end{equation}
The dichroism in absorbance depends, obviously, on the thickness of
the material. In the usual experimental assumption is that the sample
is thin and that $\Delta(\omega,\,d)$ does not depend on $d$.  As can
be seen from Fig. \ref{fig:Delta_tr}, $\Delta(\omega,\,d)$ is weakly
dependent on $d$ in the region where the two phonon resonances occur
($214.5 < \omega < 215.25$ ) only for $d\le 10^{-6}$ cm.  We note that
in this region the position of the minimum and maximum in the circular dichroic
absorption are located at the energies of phonon frequencies split by
dynamical effects.  For thicker samples, the signal is progressively
suppressed, and the minimum and maximum do not correspond anymore with
the energies of the phonon frequencies.

The limit of thickness going to zero in Eq. \ref{eq:Delta} leads to the  expression
\begin{equation}
\label{eq:Delta2}
\Delta(\omega)=\frac{{\rm Re} (\varepsilon_{xy}-\varepsilon_{yx})}{{\rm Im} (\varepsilon_{xx}+\varepsilon_{yy})}\,.
\end{equation}
that is also plotted in Fig. \ref{fig:Delta_tr} and matches very well the smallest sample thicknesses.

\subsection{Faraday rotation }

In this section, for completeness, we obtain the equation for Faraday rotation, i.e. the rotation of the polarization of light when light is transmitted through a magnetic material or in the presence of an external magnetic field.

We consider a linear polarized incident light.
The equations for the electric fields are:
\begin{align}
\label{eq:E_H_i_r_m_t_linear}
{\bf E}_{i,\theta} =&\mathbf{\hat{e}}_\theta\,e^{i\omega z/c}e^{-i\omega t}
\nonumber \\
{\bf E}_{r,\theta} =& \left(\mathbf{\hat{e}}_{x}\,r_{x,s}+\mathbf{\hat{e}_{y}}\,r_{y,s}\right) \,e^{-i\omega z/c}e^{-i\omega t}
\nonumber \\
{\bf E}_{m,\theta} =& {\bf E}_1\left(m_{11,\theta} \,e^{i\omega n_1\,z/c}+m_{12,\theta} \,e^{-i\omega n_1\,z/c}\right)e^{-i\omega t}
\nonumber \\
&+ {\bf E}_2\left(m_{21,\theta} \,e^{i\omega n_2\,z/c}+m_{22,\theta} \,e^{-i\omega n_2\,z/c}\right)e^{-i\omega t}
\nonumber \\
{\bf E}_{t,\theta} =& \left(\mathbf{\hat{e}}_{x}\,t_{x,\theta}+\mathbf{\hat{e}}_{y}\,t_{y,\theta}\right) \,e^{i\omega z/c}e^{-i\omega t}
\end{align}

The boundary conditions at $z=0$ and $z=d$ give
\begin{align}
\label{eq:bcs_linear2}
&\mathbf{\hat{e}}_\theta+\mathbf{\hat{e}}_{x}\,r_{x,\theta}+\mathbf{\hat{e}}_{y}\,r_{y,\theta}
\nonumber \\
&={\bf E}_1\,(m_{11,\theta}+m_{12,\theta})+{\bf E}_2\,(m_{21,\theta}+m_{22,\theta})\,,
\nonumber \\
&\mathbf{\hat{e}}_\theta-\mathbf{\hat{e}}_{x}\,r_{x,\theta}-\mathbf{\hat{e}}_{y}\,r_{y,\theta}
\nonumber \\
&=n_1\,{\bf E}_1\,(m_{11,\theta}-m_{12,\theta})+n_2\,{\bf E}_2\,(m_{21,\theta}-m_{22,\theta})
\end{align}
and
\begin{align}
\label{eq:bcs_linear3}
& {\bf E}_1\left(m_{11,\theta} \,e^{i\omega n_1\,d/c}+m_{12,\theta} \,e^{-i\omega n_1\,d/c}\right)
\nonumber \\
&+ {\bf E}_2\left(m_{21,\theta} \,e^{i\omega n_2\,d/c}+m_{22,\theta} \,e^{-i\omega n_2\,d/c}\right)
\nonumber \\
&= \left(\mathbf{\hat{e}}_{x}\,t_{x,\theta}+\mathbf{\hat{e}}_{y}\,t_{y,\theta}\right) \,e^{i\omega d/c}\,,
\nonumber \\
& n_1\,{\bf E}_1\left(m_{11,\theta} \,e^{i\omega n_1\,d/c}-m_{12,\theta} \,e^{-i\omega n_1\,d/c}\right)
\nonumber \\
&+ n_2\,{\bf E}_2\left(m_{21,\theta} \,e^{i\omega n_2\,d/c}-m_{22,\theta} \,e^{-i\omega n_2\,d/c}\right)
\nonumber \\
&= \left(\mathbf{\hat{e}}_{x}\,t_{x,\theta}+\mathbf{\hat{e}}_{y}\,t_{y,\theta}\right) \,e^{i\omega d/c}\,,
\end{align}
respectively. We now have eight unknowns, $r_{x,\theta}$, $r_{y,\theta}$, $m_{11,\theta}$, $m_{12,\theta}$, $m_{21,\theta}$, $m_{22,\theta}$, $t_{x,\theta}$, and $t_{y,\theta}$, and eight equations [Eqs.~\eqref{eq:bcs_linear2} and~\eqref{eq:bcs_linear3}].

The polarization angle and the ellipticity for transmitted light can be obtained using Eqs.~\eqref{eq:Stokes_parameters}, \eqref{eq:pol_direction}, and~\eqref{eq:ellipticity}, except that now  $t_{x,\theta}$ and $t_{y,\theta}$ must be used at places of $r_{x,\theta}$ and $r_{y,\theta}$, respectively.

\subsection{Effect of a large drag-force coefficient.}

In the calculations of the magneto-optical effects carried out in the
previous sections, we consider a drag force coefficient
$\tilde{\Gamma}=0.02$ cm$^{-1}$ corresponding to the case of
$\tilde{\Gamma}\ll \Delta \omega=0.22$ cm$^{-1}$, where
$\Delta \omega$ is the non-adiabatic phonon splitting calculated
within time dependent density functional theory in the adiabatic
local-density approximation. This approximation assumes, however,
adiabatic separation between the ionic and spin dynamics and can
result in incorrect values of the non-adiabatic phonon
splitting. Thus, in the experimental situation, the phonon splitting
could even be smaller than $\tilde{\Gamma}$.  It is then important to
verify how the calculated spectra depends on the ratio
$\tilde{\Gamma}/\Delta \omega$ to verify that the circular dichroic
signal remains visible even under this condition.
\begin{figure}
\includegraphics[width=1.1\columnwidth]{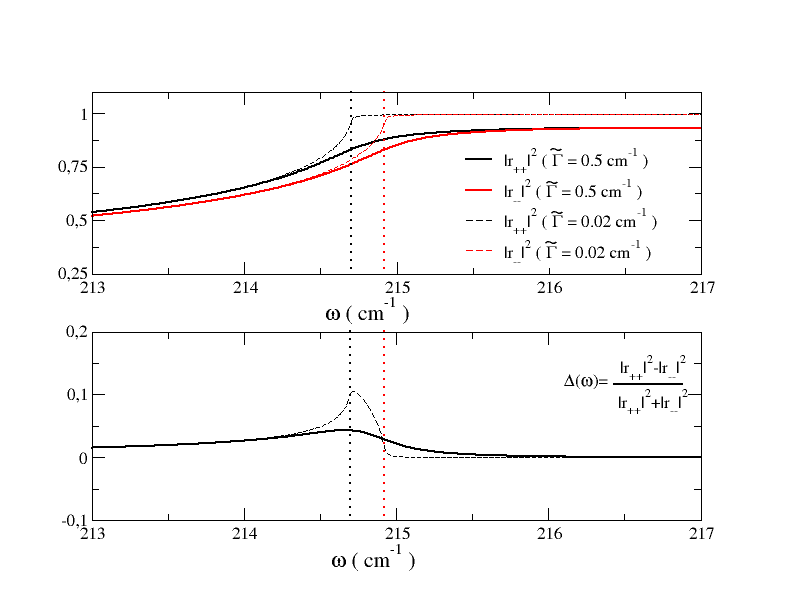}
    \caption{Reflectivity (top) and Dichroic absorption (bottom) in  CrI$_3$ in the assumption of a measurement in reflection from an infinite sample and for two different values of the drag force coefficient. The vertical dotted lines mark the energy of the infrared frequencies split by dynamical effects. }    \label{fig:Reflection_large_res}
\end{figure}

In Fig. \ref{fig:Reflection_large_res} we plot the left and right
reflectivities in reflection from an infinite sample and for a drag
force coefficient that is larger than twice the splitting of the
phonon modes. As can be seen in the bottom panel, the circular
dichroic signal is still clearly visible.

\begin{figure}
\includegraphics[width=1.1\columnwidth]{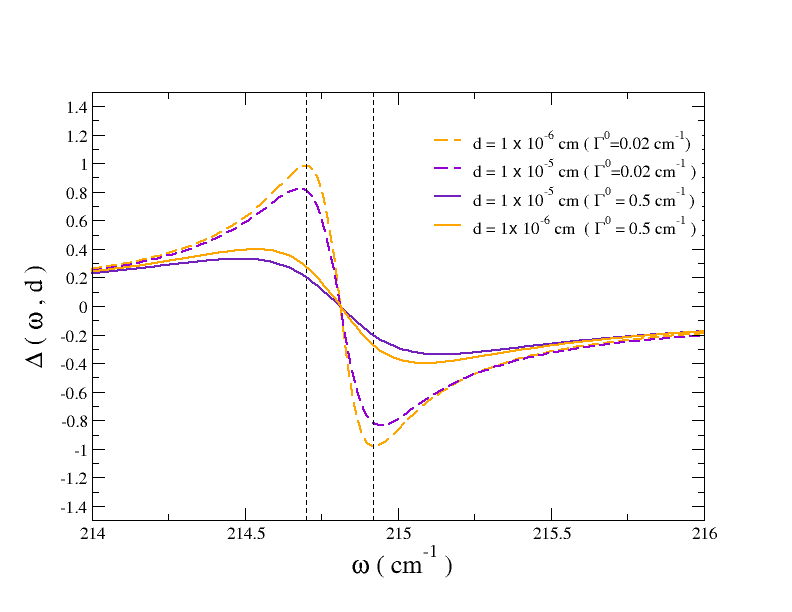}
    \caption{Dichroic absorption in  CrI$_3$ in the assumption of a measurement in transmission from a sample of thickness $d$ as a function of the force drag coefficient $\tilde{\Gamma}$.
    The vertical dashed lines mark the energy of the infrared phonons split by dynamical effects.  }    \label{fig:Delta_transmission_large_Res}
\end{figure}

The same occurs for the circular dichroism in transmission (see Fig. \ref{fig:Delta_transmission_large_Res}). The larger force drag coefficient reduces the amplitude of the $\Delta(\omega,d)$ signal that, even in this case, remains detectable. However, for large $\tilde{\Gamma}$, the minimum and maximum of the circular dichroic signal do not measure directly the energy of the split infrared phonon frequencies.

\section{Conclusions}

Probing phonon chirality in solids is not easy as the the phonon spectrum splitting induced by non-adiabatic effects  are generally small despite the value of the phonon angular momentum comparable to the spin one\cite{Zhang_NiuPhysRevLett.112.085503}.

In our work, we have shown that a reliable probe of phonon chirality
is the measurement of circular-dichroic infrared absorption.  Indeed,
a finite circular dichroic infrared absorption arises solely from
dynamical effects and it is exactly zero in the static case.

We obtain the equations for the optical absorption and the Faraday and Kerr effects. 
We applied our derivation to the case of CrI$_3$ and have shown that
the circular dichroic absorption  is non-negligible and sizeable, both in reflection and in transition. 

Finally, we point out that although our calculation suffers from the neglecting of the resonance with the magnon modes, resulting in an incorrect phonon splitting, we believe it to be quite reliable and significant for what concerns the intensity of the circular dichroic signal. What we expect is mostly a change in the magnitude of the splitting of the infrared active phonons, that in our case is probably too small.

{\it Note added:} In the final stages of this work,
we became aware of a preprint~\cite{Royo_Stengel_Arxiv} calculating the infrared circular dichroic absorption both with frozen spins (similar to the present work), and with relaxed spins (including the resonance with the magnons). In the frozen-spin case, the authors find qualitatively similar results to ours. Moreover, they obtain a strong enhancement of the circular dichroic signal when spin-canting is allowed, confirming the importance of the effect.

\acknowledgements
We acknowledge the useful discussions with O. Bistoni, M. Furci, P. Giannozzi and M. Stengel. 
C.-H.P. was supported by the Korean NRF No-2023R1A2C1007297. I.~S. was supported by Grant No.  PID2021-129035NB-I00 funded
by MCIN/AEI/10.13039/501100011033 and by ERDF/EU. 
M.C  was funded by the European Union (ERC, DELIGHT, 101052708). Views and opinions expressed are however those of the author(s) only and do not necessarily reflect those of the European Union or the European Research Council. Neither the European Union nor the granting authority can be held responsible for them. 
\bibliography{bibliography}
\end{document}